
\input harvmac
%
%
%
%
%

\parindent=0pt
\Title{SHEP 95-07}{Non-Compact Pure Gauge QED in 3D is Free}

\centerline{\bf Tim R. Morris}
\vskip .12in plus .02in
\centerline{\it Physics Department}
\centerline{\it University of Southampton}
\centerline{\it Southampton, SO17 1BJ, UK}
\vskip .7in plus .35in

\centerline{\bf Abstract}
\smallskip 
For all Poincar\'e invariant Lagrangians of the form ${\cal L}\equiv
f(F_{\mu\nu})$, in three Euclidean dimensions, where $f$
is any invariant function of  a non-compact
$U(1)$ field strength  $F_{\mu\nu}$,
we find that the only continuum limit (described by just such a gauge field)
is that of
free field theory:  First we approximate
a gauge invariant version of Wilson's renormalization
group by neglecting  all higher derivative terms $\sim \partial^nF$
in ${\cal L}$, but allowing for a general non-vanishing anomalous dimension.
Then we prove analytically that the resulting flow equation has only
one acceptable fixed point: the Gaussian fixed point.
The possible relevance to high-$T_c$ superconductivity is briefly discussed.

\vskip -1.5cm
\Date{\vbox{
{hep-th/9503225}
\vskip2pt{March, 1995.}
}
}
\catcode`@=11 
\def\slash#1{\mathord{\mathpalette\c@ncel#1}}
 \def\c@ncel#1#2{\ooalign{$\hfil#1\mkern1mu/\hfil$\crcr$#1#2$}}
\def\lsim{\mathrel{\mathpalette\@versim<}}
\def\gsim{\mathrel{\mathpalette\@versim>}}
 \def\@versim#1#2{\lower0.2ex\vbox{\baselineskip\z@skip\lineskip\z@skip
       \lineskiplimit\z@\ialign{$\m@th#1\hfil##$\crcr#2\crcr\sim\crcr}}}
\catcode`@=12 
\def\nonp{non-perturbative}
\def\phi{\varphi}
\def\th{\vartheta}
\def\te#1{\theta_\epsilon( #1,\Lambda)}
\def\epsilon{\varepsilon}

\def\q{{\bf q}}

\def\tr{{\rm tr}}
\def\D{{\cal D}}
\def\E{{\cal E}}
\def\G{{\cal G}}
\def\H{{\cal H}}
\def\L{{\cal L}}
\def\ins#1#2#3{\hskip #1cm \hbox{#3}\hskip #2cm}
\def\frac#1#2{{#1\over#2}}
\def\Fe{F_{\mu\nu}}
\def\d#1{{\rm dim}(#1)}
\parindent=15pt

{}From the point of view of perturbation theory, the question of whether there
are any non-trivial continuum limits (in other words renormalizable
interacting field theories) of just a single $U(1)$ gauge field $A_\mu$,
seems absurd. After all, the canonical mass dimension of the gauge
invariant field strength $\Fe$ is $D/2$, in $D$ dimensions, and thus
the simplest gauge invariant scalar combination\foot{We will discuss
a Chern Simons term, possible in $D=3$ dimensions, at the end.}\
$\Fe\Fe$ is already of dimension $D$,
and all other gauge invariant scalar combinations will be non-renormalizable,
since they have dimension larger than $D$. In other words all gauge invariant
interactions will be irrelevant and only the free theory ${\cal L}\sim\Fe\Fe$
is left once the ultra-violet cutoff is removed.

However, this argument is
only valid in the perturbative regime. Non-perturbatively it can happen that
na\"\i vely irrelevant operators, by receiving large anomalous dimensions,
are actually marginal or relevant. (This happens, for example, to the
four-fermi
coupling in the apparent strong coupling continuum limit of four dimensional
QED\ref\bill{C.N. Leung, S.T. Love and W.A. Bardeen, Nucl. Phys. B273 (1986)
649\semi M. Gockeler et al, Nucl. Phys. B371 (1992) 731.}).
In fact, the  {\sl compact} $U(1)$  version of lattice pure gauge QED is far
from trivial in three
dimensions, giving a
confined disordered phase resulting from monopole
condensation\ref\polyi{ A. M. Polyakov,
Phys. Lett. 59B (1975) 82}.  The difference between compact and non-compact
$U(1)$ gauge theory lies in whether, in a lattice formulation,
the $U(1)$ gauge transformations (and correspondingly the
bare connections $A_\mu$) are valued on a circle or the real line. In
the continuum this
translates into whether
monopole field configurations are in principle allowed or not.
Here we will
be working with non-compact QED. We intend to discuss the compact case in
a separate publication.

Notice that {\sl if} there exists a non-trivial continuum limit for pure
gauge non-compact QED,
then it cannot be reached from a bare Lagrangian formulated
about the above Gaussian fixed point, since this is I.R. attractive.
In other words, the theory must be strongly interacting also at the cutoff
scale $\Lambda_0$. In this case we do not know {\it a priori} what form
to take for the (local) bare Lagrangian, and indeed there is no reason to
assume
that it is even polynomial in the fields. For this reason we must start
with as general a local Lagrangian as possible.

One main motivation for this letter is the continuing speculation
that some sort of strongly coupled fixed point
involving a dynamically
generated $U(1)$
gauge field could be responsible for  high-$T_c$
superconductivity\ref\wilc{
G. Baskaran and P.W. Anderson, Phys. Rev. B37  (1988) 580;\quad
P.A. Lee, Phys. Rev. Lett. 63 (1989) 680;\quad
L.B. Ioffe and A.I. Larkin, Phys. Rev. B39 (1989) 8988;\quad
B. Blok and H. Monien, Phys. Rev. B47 (1993) 3454;\quad
J. Gan and E. Wong, Phys. Rev. Lett. 71 (1993) 4226;\quad
C. Nayak and F. Wilczek, Nucl. Phys. B417 (1994) 359;\quad
S. Khlebnikov, Phys. Rev. B50 (1994) 6954;\quad
J. Polchinski, Nucl. Phys. B422 (1994) 617;\quad
S. Chakravarty, R.E. Norton and O.F. Sylju\aa sen, Phys. Rev. Lett. 74
(1995) 1423;\quad
B.L. Altshuler, L.B. Ioffe and A.J. Millis, Phys. Rev. B50 (1994) 14048.
 }.
In this case also, there is no {\it a priori} reason to restrict the bare
phenomenological Landau Ginzburg Lagrangian to quadratic in the $U(1)$
gauge field, since the gauge field is strongly interacting at the lattice
level. A strongly coupled fixed point for the pure gauge sector could
conceivably control the dynamics of the (massless) gauge field at
energy scales
much lower than the masses of all the other quasiparticles, or indeed
to an extent at energy scales above these excitations
if the pure gauge sector is still close to this fixed point.
However,
as already stated in the abstract, we shall find that even for a Lagrangian
consisting of the most general function of the field strength, and allowing
for any anomalous dimension for $A_\mu$, the only fixed point is
the trivial Gaussian one -- thus ruling out 
any fundamental
non-linear generalisation of pure gauge QED in three dimensions.
 This only indicates that if such a
fixed point exists, then it cannot be realised in a three dimensional
Poincar\'e invariant non-compact local theory without the inclusion of
other dynamical fields.\foot{Of course, it is the restriction to {\sl local}
effective
actions that disallows other propagating
low energy excitations from already being
hidden in poles and cuts of the effective vertices.}
Nevertheless, we feel it is worthwhile to emphasise the
possibility that the low energy excitations might be described by a
phenomenological (continuum) theory whose {\sl bare} action is
{\sl not defined about
a Gaussian  Ultra Violet} fixed point.

We make the approximation of dropping
all momentum dependence in the effective Lagrangian,
and correspondingly in the renormalization group flow,
beyond that contained in a general function of
the field strength. Nevertheless, this is
already sufficient to allow for general wavefunction renormalization --
and such approximations have so far proved very\nref\hasetal{
F.J. Wegner and A. Houghton, Phys. Rev. A8 (1973) 401\semi
A. Parola and L. Reatto, Phys. Rev. Lett. 53 (1984) 2417\semi
A. Parola, J. Phys. C19 (1986) 5071\semi
A. Hasenfratz and  P. Hasenfratz, Nucl. Phys. B270 (1986) 685\semi
G. Felder, Comm. Math. Phys. 111 (1987) 101.}\
robust\ref\deriv{T.R. Morris, Phys. Lett. B329 (1994) 241.}\ref\twod{T.R.
Morris, Phys. Lett. B345 (1995) 139.}, in the sense that one finds
all, and only, the continuum limits expected and these are described with a
fair accuracy.
The sequence of
two dimensional multicritical examples in the latter reference
is particularly
significant, since their description is well outside the capabilities of
other approximate methods. Also approximations where
only a general potential
for the field is
kept\hasetal--\ref\trunc{T.R. Morris, Phys. Lett. B334 (1994) 355.}
 have in the same sense proved robust, only failing to find
 the two dimensional
multicritical examples where the fact that this further approximation
sets anomalous dimensions to zero, restricts qualitatively the allowed
continuum limits. (A
scalar field in two dimensions has vanishing canonical dimension
so that, by scaling, power law behaviour for large field is ruled out
in this approximation\twod).
Therefore we  believe that our 
conclusion
is correct also for the exact theory.

Our second main motivation is to apply  these methods of approximation to a
gauge invariant system in as simple a setting as possible.
It must be emphasised that the problems
posed by gauge invariance, in these methods, are apparently not ones of
principle but of practice: to make
the approximations manageable it is most convenient to place the cutoff
in a free (inverse) propagator -- but this typically breaks the gauge
invariance, with the consequence that BRST invariance
has to be imposed by hand (on renormalised quantities),
and can only be exactly satisfied once
the cutoff is removed.\foot{A background gauge invariant method
is proposed in ref.\ref\wet{M. Reuter and C. Wetterich, Nucl. Phys.
B408 (1993) 91; B417 (1994) 181; B427 (1994) 291.}, but the crucial
problem of
broken BRST invariance is not addressed.}
 Although perturbation theory has been succesfully
addressed\ref\pertg{B. Warr, Ann. Phys. 183 (1988) 1 and 59\semi
G. Keller and C. Kopper, Phys. Lett. B273 (1991) 323\semi
C. Becchi, in ``Elementary particles, field theory and statistical mechanics'',
Eds. M. Bonini, G. Marchesini and E. Onofri, (1993) Parma University\semi
M. Bonini, M. D'Attanasio and G. Marchesini,  Nucl. Phys. B418 (1994) 81;
B347 (1995) 163; Parma preprint (1994), UPRF-94-414, hep-th/9412195.}, the
methods do not easily generalise to workable \nonp\ approximations\ref\ell{U.
Ellwanger, Phys. Lett. B335 (1994) 364;\quad
U. Ellwanger, M. Hirsch and A. Weber, preprint LPTHE Orsay 95-39,
hep-ph/9506019.}.
In this letter we effectively sidestep these issues by concentrating on
pure $U(1)$ gauge theory.

Actually, there is a possibly greater technical challenge:
the method becomes increasingly more difficult
to use systematically, as the number of invariants grows.
This is because the flow equations are expressed as non-linear
partial differential equations in the scale and each independent invariant,
which then generally have to be solved numerically.
So far, only systems with functions of {\sl one
invariant} 
have
been considered without further approximation.
As shown in appendix A, an $SU(N)$ field
strength $F^a_{\mu\nu}$ has   ${1\over2}(D^2-D-2)(N^2-2)-1$
invariants (if $D>2$),
which means that one has for example,
a partial differential equation in 34 invariants for pure glue
QCD at the lowest order of the derivative expansion. If we want to restrict
the discussion to just one invariant
then we are limited to
two dimensional $SU(2)$ Yang-Mills, or three dimensional $U(1)$
gauge theory.

Because we will only
consider the case of pure $U(1)$, i.e. without fermions,
it is easy to preserve gauge invariance: the point
is that all propagators can couple only to field strengths $F_{\mu\nu}$
which are transverse for all momenta (as opposed to currents
which are generally only transverse on-shell),
so it is completely irrelevant whether
we gauge fix or not. (We will return to this point at the end). It is only
necessary to couple the cutoff only to field strengths, and to
introduce a source that is also explicitly gauge invariant,
i.e. a term of the form $J_\mu A_\mu$ where $J_\mu$ is transverse. Rather
than carrying around this constraint on $J_\mu$, we solve it by
replacing $J_\mu\mapsto P_{\mu\nu}J_\nu$. Here $P_{\mu\nu}=\delta_{\mu\nu}
-{\partial_\mu\partial_\nu\over\lform}$ is the projector onto the transverse
space. Thus, following refs.\deriv\trunc, we take for the partition function
\eqn\ZZ{\exp W[J] =\int\!\D A\
\exp\{-{1\over4}F_{\mu\nu}.C^{-1}.F_{\mu\nu}
-S_{\Lambda_0}[F_{\mu\nu}]+J_\mu.P_{\mu\nu}A_{\nu}\}\ \ .}
The additive infrared cutoff will be taken to
be $C^{-1}(q,\Lambda)=1/\te q -1$, where $\te q$ is smooth and
satisfies $0<\te q<1$ for all (positive) $\Lambda$ and
$q$,  but $\te q\to \theta(q-\Lambda)$ as $\epsilon\to0$.
We use a sharp cutoff because the flow equation is
simpler, even though it does not allow an analytic momentum
expansion, but only an expansion in homogeneous
functions of momenta of integer degree $\sim p^m$\ref\erg{T.R. Morris,
Int. J. Mod. Phys. A9 (1994) 2411.}\ref\oneday{
T.R. Morris, Southampton preprint SHEP 95-21 in preparation.}. We
expect that a similar computation can be worked through with a smooth
cutoff, but the lowest order
sharp cutoff equations are just as
robust\hasetal\trunc\ as those obtained with smooth cutoffs\deriv\twod,
so we expect the conclusions to remain unchanged. Using the
fact that $P_{\mu\nu}P_{\nu\sigma}=P_{\mu\sigma}$, we have
$${\partial\over\partial\Lambda}W[J]=
{1\over2}\left\{ {\delta W\over
\delta J_\mu}.\lform{\partial C^{-1} \over\partial\Lambda}.{\delta
W\over\delta J_\mu} + \tr\left(\lform
{\partial C^{-1} \over\partial\Lambda}.{\delta^2
W\over\delta J_\mu\delta J_\mu}\right)\right\}\quad .$$
{}From now on we will suppress Lorentz indices where contractions are clear.
We transform to the Legendre effective action by writing
$\Gamma[A]-{1\over2} A.(\lform C^{-1} P).A =-W[J]+J.P.A$, where
$P.A=\delta W/\delta J$ and $\delta \Gamma/\delta A -\lform C^{-1} P.A=
P.J$. From these latter relations, it follows that
$P.{\delta\over\delta J}={\delta\over\delta J}$ and
$P.{\delta\over\delta A}={\delta\over\delta A}$
and hence,
$${\partial\over\partial\Lambda}\Gamma[A]=
-{1\over2}\tr\left[{P\over C}{\partial C\over \partial\Lambda}
.\left(1-{C\over\lform}.
{\delta^2\Gamma\over\delta A\delta A}\right)^{-1}\right]
\ \ ,$$
where the inverse is defined in the transverse space.
 As discussed above,  to lowest order we can write, in three dimensions,
$\Gamma[A]={1\over4}\int\! d^3x\, \L(F^2_{\mu\nu},\Lambda)$,
for some function $\L$. Since we are dropping all space-time derivatives
of $F$, we have
${\delta^2\Gamma\big/\delta A_\lambda\delta A_\sigma}\equiv
-4 \L''F_{\mu\lambda} F_{\alpha\sigma}\partial_\mu\partial_\alpha
-\L'(\lform\delta_{\lambda\sigma}-\partial_\lambda\partial_\sigma)$,
where primes refer to derivatives with respect to $F^2_{\mu\nu}$.
Adapting from ref.\deriv, we thus have
$$\eqalign{&\left[1-
{C\over\lform}.{\delta^2\Gamma\over\delta A\delta A}\right]^{-1}
\mkern-23mu(\q,-\q)_{\mu\nu}P_{\mu\nu}(q)=\cr
&\int\! d^3x\, \left\{ {1\over1+\L'C(q,\Lambda)}+{1\over1+[\L'+4\L''
F_{\mu\lambda}F_{\mu\sigma}q_\lambda q_\sigma/q^2]C(q,\Lambda)}\right\}\ ,}$$
and hence, rotating $q_{\mu}\mapsto R_{\mu\nu}q_{\nu}$
so that $F_{\lambda\sigma}
R_{\lambda\mu} R_{\sigma\nu}=\epsilon_{\mu\nu3}\sqrt{\half F^2_{\alpha\beta}}$,
we have
$$\eqalign{{\partial\Gamma[A]\over\partial\Lambda}&=
-{1\over(2\pi)^2}\int\!d^3x\int_0^\infty\!\! dq\, {q^2\over\te q}
{\partial\te q\over\partial\Lambda}\int^\pi_0\!{d\th\over2}\sin\th\cr
&\left\{ {1\over 1+\te q(\L'-1)}+{1\over1+\te q(\L'-1+2\L''F^2\sin^2\!\th)}
\right\}\ .}$$
Now we take the limit $\epsilon\to0$ using the relation \trunc\erg:
$${1\over\te q}{\partial\te q\over\partial\Lambda}{1\over
1+\te q f(q,\Lambda)}
\to \delta(q-\Lambda)\ln[1+f(q,\Lambda)] + \hbox{const.}\ \ ,$$
where $f$ is any smooth function. The (infinite) constant yields a field
independent vacuum energy
which can be adsorbed by a shift in $\L$. The $q$ integral is then trivial.
We perform the $\th$ integral, and change to dimensionless (renormalised)
variables so that  $t=\ln(\Lambda_0/\Lambda)$,
$F^2=\zeta\Lambda^{3+\eta}\phi/\Lambda_0^\eta$
and $\L(F^2,\Lambda)
\mapsto \zeta \Lambda^3
\left[\L(\phi,t) - \eta\ln(\Lambda_0/\Lambda)
 - \eta/3\right]$, where $\eta$ is the
conventional anomalous dimension: $[A_\mu]=\half(1+\eta)$, and
$\zeta=2\big/3\pi^2$. The result is
\eqn\flo{
{\partial\over\partial t}\L(\phi,t)+\left(1+{\eta\over3}\right)\phi\L'-\L=
P\left({2\phi\L''\over\L'}\right)+\ln\L'-1\quad,}
where prime now refers to differentiation with respect to $\phi$, and
$$\eqalign{
P(w) &=
\sqrt{1+w\over w}\tanh^{-1}\sqrt{w\over1+w}\ins11{if} w>0\cr
 &=\sqrt{1+w\over -w}\tan^{-1}\sqrt{-w\over1+w}\ \ \ins11{if} -1<w<0\ \ ,
\cr}$$
and $\tan^{-1}$ is taken in the range $0\le \tan^{-1}\le \pi/2$.
The flow equation
holds true only if the physical stability requirements
\eqn\sta{\L'>0 \ins22{and} \L'+2\phi\L''>0}
are satisfied, for otherwise, for all $\epsilon>0$,
 the $q$ integral diverges at unphysical poles.
Note that $\phi$, being a rescaled version of $F^2$, only has physical
meaning for $\phi\ge0$.

Finally, from \flo, all
massless continuum limits (i.e. fixed points  $\partial\L/\partial t=0$)
satisfy
\eqn\fp{\left(1+{\eta\over3}\right)\phi\L'(\phi)-\L(\phi)=
P\left({2\phi\L''(\phi)\over\L'(\phi)}\right)+\ln\L'(\phi)-1\quad.}
All massive continuum limits result from the tuning of relevant and
marginal couplings,
as such a fixed point is approached in the limit $t\to\infty$.

Equation \fp\ has at least one solution, namely the
Gaussian fixed point:
\eqn\gau{\L(\phi)={\rm e}^{-\E}\phi+\E\ins22{and}\eta=0\quad.}
The value of the real constant $\E$ here is quite irrelevant
(e.g. choose $\E=0$),
because the approximation preserves a field reparametrization
invariance:
\eqn\inv{\phi\mapsto\lambda\phi\ins11,\L\mapsto\L-\ln\lambda\quad,}
as it must if $\eta$ is to be determined\deriv\ref\revi{T.R. Morris, in {\it
Lattice '94}, Nucl. Phys. B(Proc. Suppl.)42 (1995) 811.}\ref\geoff{The problems
that arise when field
reparametrization invariance is broken are discussed in:\quad
T.L. Bell and K.G. Wilson, Phys. Rev. B11
(1975) 3431\semi
E.K. Riedel, G.R. Golner and K.E. Newman, Ann. Phys. 161 (1985) 178\semi
G.R. Golner, Phys. Rev. B33 (1986) 7863.}.
Let us briefly adapt those arguments\deriv\revi\ in order to show that at
 most a countable number of acceptable
solutions are expected from  \fp, 
before going on to prove that \gau\ is the
only one.

The central assumption is that any acceptable
Lagrangian $\L$ must be well defined for all values of $\phi\ge0$.
If $\L(\phi)$ is regular as $\phi\to0$, then because $\L''$ drops out of \fp\
in that limit, the solution for $\L$ contains only one free parameter, (for
given $\eta$):
\eqn\zer{\L(\phi)=\E+{\rm e}^{-\E}\phi+
{\eta\over10}{\rm e}^{-2\E}\phi^2+O(\phi^3)\quad.}
On the other hand if $\L$ is well defined for all $\phi\ge0$, then
it is easy to convince oneself that  for $\phi\to\infty$, $\L$ must behave
as
\eqn\inf{\L(\phi)=A\phi^{3/(3+\eta)}+{\eta\over3+\eta}\ln\phi+\cdots}
[the dots are a calculable constant and $O(\phi^{-6/(3+\eta)})$], i.e.
 to leading order according to the scaling dimensions of $\L$ and $\phi$.
Since this latter equation also contains only one free parameter, i.e. $A$,
\zer\ and \inf\ provide sufficient boundary conditions to allow at most
a discrete set of solutions to \fp, for given  choice of $\eta$. However
the reparametrization invariance \inv\ provides an extra constraint, since
it implies that e.g. \zer, is already sufficient to determine the
equivalence class of solutions [under \inv] uniquely.
Thus the invariance \inv\ leads to an overconstrained solution space
and results in quantization of $\eta$.
In this way, the fixed point equation
\fp\ may be regarded as a non-linear eigenvalue equation for $\eta$.

(We mention briefly the results one obtains from truncations to
 polynomial field
dependence: $\L\equiv\E+\sum_{m=1}^M a_m ({\rm e}^{-\E}\phi)^m$. This amounts
to declaring $a_{M+1}=0$. The $a_m$ turn out to be polynomials
of $\eta$ with positive coefficients, e.g. 
$a_3=(101\eta+
75)\eta/5250$ and $a_4=(3746\eta^2+6350\eta+1875)\eta/787500$, whose vanishing
yields of course the Gaussian solution $\eta=0$, but also real negative
solutions for $\eta$: for $M=2$, $\eta=-75/101$; for $M=3$,
$\eta=-1.31, -.380$; for $M=4$, $\eta=-1.66,-.852, -.204$; etc.
Nevertheless all these  `approximate' non-zero solutions are
completely spurious, which fact
serves to reemphasise the unreliability of truncations\trunc.)

What happens to the solutions at the `wrong' values of $\eta$?
These solutions  do not make it  out to $\phi=\infty$, but instead die
in one of two ways, at some finite 
$\phi=\phi_c$:\eqna\si
$$\eqalignno{
\L =&{3\over2(3+\eta)}x\left\{\ln x-\ln(-\ln x)\right\}+\cdots &\si a\cr
\rm{or}\quad\L =&1-\ln\left({1-\pi c\over(1+\eta)\phi_c}\right)
+{1-\pi c\over1+\eta}\left\{ 1+{\eta\over3}
-x+{x^2\over4}+{x^3\over24}+{3+8c^2\over192}x^4+O(x^5)\right\}\cr
&{} &\si b\cr
}$$
where $x=1-\phi/\phi_c$. In type \si{a}, the dots refer to less singular,
and non-singular, terms. In type \si{b}, the constant $c>0$ for the
solution to be valid for $\phi<\phi_c$, and the solution is chosen
so that $P(2\phi\L''/
\L')=\pi c x/2+O(x^2)$. It is not intrinsically singular, but
satisfies $\L'+2\phi\L''=0$ at $\phi=\phi_c$, violating the stability
conditions \sta\ here, and for $\phi>\phi_c$ it no longer satisfies \fp,
but an analytic continuation of the fixed point equation
where $P$ is replaced by $-P$. Note that in common with previous findings
in scalar field theory\deriv--\trunc, the `wrong' Lagrangians do
not {\sl diverge} as $\phi\to\phi_c^-$ (which it might be argued could be
physically acceptable).

We now prove that the only non-singular solution of fixed point equation
\fp\ is the Gaussian fixed point \gau.
 First we
recognize that the reparametrization invariance \inv\ allows us to convert
\fp\ into an autonomous (viz. translation invariant)
 second order ordinary differential equation. Thus
if we define $z=\ln\phi$ and $U(z)=\L(\phi)-z$, then \fp\ becomes
\eqn\Ueq{
\left(1+{\eta\over3}\right)(1+U')-U=P\left(2\left[{U''\over1+U'}-1\right]
\right)+\ln(1+U')-1\quad,}
subject 
to the requirements \sta: $2U''>1+U'>0$,
which in particular imply
\eqn\bou{[\ln(1+U')]'>1/2\quad.}
Integrating this inequality we see that any non-singular solution
of the fixed point equation has the property that
$U'(z)$ is monotonic increasing, and
passes through zero. (And in fact obeys
$U'(z)\to -1$ as $z\to-\infty$, and $U'(z)\to\infty$
as $z\to\infty$).  We use the translation invariance
of \Ueq\ to set  $U'(0)=0$. In other words we have that such a $U(z)$ may be
taken to be a
decreasing function for all $z<0$, with a minimum at $z=0$, and increasing
thereafter.  Moreover, we arrive at
\proclaim Lemma (i). Any non-singular solution
$U(z)$ is unbounded from above both in the region
$\{ z : z<0\}$ and in the region $\{z : z>0\}$.

Now we proceed by assuming
 that the solution is non-singular, and show that this contradicts
lemma (i), for all but the Gaussian fixed point. In terms of
$Y(U)=1+U'-\ln(1+U')$,
the fixed point equation \Ueq\
becomes first order:
\eqn\Yeq{P\left(2[dY/dU-1]\right)={\eta\over3}\left(1+U'\right)
+Y-U+1\quad,}
where we have used $dY/dU = U''/(1+U')$.
Note that, for fixed sign of $z$,
$Y(U)$ is a single valued function
and  $U'$ may be regarded as a single valued function of $Y$.
It will be useful also to note that, from \bou, $dY/dU>1/2$.
Now we divide the analysis of the behaviour of $U(z)$
into five separate cases: $\eta>0$ and $z<0$, $\eta>0$ and $z>0$,
 $\eta<0$ and $z<0$, $\eta<0$ and $z>0$, and finally $\eta=0$ (and any $z$).

First we assume $\eta>0$, $z<0$
and $U''(0)<1$.
We note that \Yeq\ implies
\eqn\du{{\partial \over \partial U}P\left(2\left[{dY\over dU}-1\right]\right)
=-1 +\left\{1+{\eta\over3}{1+U'\over U'}\right\} {dY\over dU}\quad.}
 For all $z<0$, the factor in curly brackets is less than one.
Thus since $\lim_{z\to 0}dY/dU\equiv U''(0)<1$, and
 $P(w)$ is a monotonically
increasing function of $w$, we have for all $z<0$ that  $dY/dU<U''(0)$.
Therefore from the above equation we have that $\partial P/\partial
U<U''(0)-1$, in this region.
 But, by integrating this inequality with respect to $U$,
and using $P\ge 0$, one obtains that $U$
is bounded above, violating (i).
(Clearly this corresponds
to encountering the singular behaviour \si{b} at some $z=z_c<0$).

Next we assume $\eta>0$, $z>0$ and $U''(0)>3/(3+\eta)$. Regard $U$ as a
(single valued) function of $Y$. Differentiating \Yeq\
with respect to $Y$ gives
\eqn\dy{2P'\left(2\left[{dY\over dU}-1\right]\right){d^2U\over dY^2}=
\left({dU\over dY}\right)^2\left[{dU\over dY}-1-{\eta\over3}{1+U'\over
U'}\right]
\quad.}
Hence, since $P'(w)>0$ and non-singular for all $w>-1$,
and  $\lim_{z\to 0}dU/dY\equiv 1/U''(0) <1+\eta/3$ (and $U'>0$), we
have that $dU/dY$ is a decreasing function of $Y$, which tends to its
lower bound: $dU/dY\to0$.
Differentiating $P$, one readily derives
$P'(w)<{1\over2w}$ for all $w>0$, and hence for all $dU/dY<\zeta<1$ we have
$$-{d^2U\over dY^2}\ge2(1-\zeta)\left(1+{\eta\over3}
-\zeta\right){dU\over dY}\quad.$$
Integrating this inequality with respect to $Y$, and using $dU/dY\ge0$,
we again obtain that $U$ is bounded above, in contradiction with (i).
(Clearly this corresponds to encountering \si{a} at some $z=z_c>0$).

Thus we have found that for $\eta>0$, the solution is singular if
$U''(0)<1$ or $U''(0)>3/(3+\eta)$. But these overlapping
regions cover all possibilities for $U''(0)$, and so we conclude
that there are no non-singular solutions 
for $\eta>0$.

The remaining cases may be similarly analysed.
Consider $\eta<0$. For $z<0$ and $U''(0)>1$, we deduce from
\dy\ that $-d^2U/dY^2\ge2[1-1/U''(0)]^2dU/dY$, and hence
(i) is violated.  For $z>0$ and
$U''(0)<3/(3+\eta)$, we note that $\lim_{z\to0}dY/dU=3\zeta/(3+\eta)$
for some $\zeta<1$, and thus from \du, $\partial P/\partial U<\zeta-1$,
violating (i). Again these two regions for $U''(0)$
cover all choices, so there are no non-singular solutions for $\eta<0$.

This leaves only the possibility that $\eta=0$. Setting $\eta=0$ in \du\
we see that, for {\sl either} fixed sign of $z$, if there is some $U=U_b$
such that ${dY\over dU}(U_b)<1$, then ${\partial P\over\partial U}<
{dY\over dU}(U_b)-1$ for $U>U_b$, and hence (i) is violated.
On the other hand, setting $\eta=0$ in \dy\ we see that, again for
{\sl either} fixed sign of $z$, if there is some $Y=Y_b$ such that
${dU\over dY}(Y_b)<1$, then again (i) is violated.
 Thus we must have $dY/dU=1$ for all $U$. This implies
$U''=1+U'$ i.e. $U'+1=z+U$ (up to an arbitrary shift on $z$). But
substituting this (and $P=1$) into \Ueq\ gives
$U(z)=\e{z}-z$, that is the   Gaussian  fixed point \gau.
\vbox{\hfill$\lform$}

We finish by tying up some loose ends.
First of all, we have shown so far that
the only continuum limit with no mass parameter is the free Gaussian
field theory \gau. In principle an interacting massive theory could exist
if relevant and marginal couplings are tuned appropriately as this
fixed point is approached in the limit $t\to\infty$. But by linearising
the flow equation \flo\ about  \gau, it is
straightforward to recover (essentially)
 the standard power counting argument, mentioned at the very beginning.
Thus we find that there are no relevant or marginal operators,
except for the exactly marginal redundant
operator that generates the
invariance \inv,  and
 only free field theory results from the approach to the Gaussian
fixed point.

Secondly, the fact that we did not gauge fix the partition function \ZZ\
might have looked worrisome. Let us show that the same results would
be obtained if we had proceeded more conventionally. Thus we introduce
a gauge fixing functional linear in $A_\mu$ (e.g. $\partial_\mu A_\mu$),
and following the standard route obtain
 a gauge field propagator which is now fully invertible,
as a result of adding to $\L$ the gauge fixing Lagrangian $\xi \L_{gf}[A]$.
Now however, we note that no interactions
couple to the longitudinal part of the propagator (this is e.g.
obvious from the
expression for ${\delta^2\Gamma\big/\delta A_\lambda\delta A_\sigma}$
given earlier), so that the flow equation for $\L$ is again \flo,
independent of $\xi$.
Also, the gauge fixing Lagrangian remains unrenormalised since it is connected
by unbroken BRST invariance to the completely decoupled
free field ghost Lagrangian.

Finally, consider adding a Chern-Simons term: $\Gamma\mapsto
\Gamma+{m\over2}\int\! d^3x\,
\epsilon_{\mu\nu\lambda}\Fe A_{\lambda}$. Since this term is only gauge
invariant after integration by parts, it cannot appear multiplied by any
other terms. It adds a purely momentum dependent term
to $\delta^2\Gamma/\delta A\delta A$, and thus alters the flow equation
\flo, but because the expression for ${\delta^2\Gamma\big/\delta
A_\lambda\delta A_\sigma}$ given earlier is
 gauge invariant,
no corrections to the Chern-Simons term are generated. Indeed,
this conclusion holds to all orders of the derivative (momentum scale)
expansion.
It follows that the Chern-Simons coupling $m$ simply flows according to
its scaling dimension $[m]=1-\eta$. Thus for any $\eta\ne1$, $m$ vanishes
at a fixed point, and  we recover the same equation \fp, and
conclusions. 
For $\eta=1$, $m$ is exactly marginal,
and  $[A_\mu]=1$. Using
the invariance \inv\ and parity, we can always choose $m=1$. What has
happened is that the Chern-Simons term has taken on the r\^ole of the
scale free normalised kinetic term, with the rest
to be considered as interactions. Since these interactions now are na\"\i vely
even more irrelevant, and
with no longer the possibility of a negative anomalous
dimension for $A_\mu$, we conclude that the appearance of a
non-trivial fixed point here
 is unlikely. However, a conclusive demonstration would require showing
(probably numerically) that the modified fixed point equation for $\L$
has no non-singular solutions. (The equivalent boundary conditions to \zer\
and \inf, lead one to expect at most a discrete set of such solutions).

\bigbreak\bigskip\bigskip\centerline{{\bf Acknowledgements}}\nobreak
It is a pleasure to thank
Ian Aitchison for providing much of the motivation through discussions of
high-$T_c$, Simon Hands and Mike Teper for perspicacious comments on
lattice $U(1)$, Ken Barnes and Ron King for discussions on invariants,
and the SERC/PPARC for providing financial support through an Advanced
Fellowship.


\appendix{A}{The number of invariants.}
We count the number of independent
invariants appearing in a general invariant function $f(F)$, where
$F$ is regarded as valued in the vector space
$\H\otimes \G$, and $\H$ ($\G$) are the Lie algebras corresponding to the
simple groups $H$ ($G$).  Note that $f$
 is not a function of all $\d{H}\d{G}$ independent
components of $F$, since it is constrained to satisfy the
invariance conditions
 $([h,F], {\partial f/\partial F})=0$, where $h$ is in the Lie algebra
of $H\otimes G$.
 Thus the number of invariants is given by $\d{H}\d{G}-\d{H}-\d{G}+\d{\Sigma}$,
where $\Sigma$ is the {\sl minimal}
 little group\ref\oraf{See e.g. L. O'Raifeartaigh, ``Group structure of gauge
theories'', (1986) CUP.}\ formed from generators $h$ that commute with $F$.
If $H$ (respec. $G$)
is dimension $1$, then $\d{\Sigma}$ is clearly the rank of $G$ (respec. $H$),
otherwise it is easy to convince oneself that $\d{\Sigma}=0$.
The formulae in the letter follow from identifying $H$ with $O(D)$, and
$G$ with $SU(N)$ or $U(1)$.

\listrefs

\end